\documentclass[aps,prx,twocolumn,superscriptaddress,]{revtex4-2}
\pdfoutput=1    

\usepackage{mathtools}
\usepackage[hidelinks,pdfusetitle,]{hyperref}

\def\tdpdf/{3D-$\Delta$PDF}
\def\hbco/{HgBa$_2$CuO$_{4+\delta}$}

\begin{document}

\title{Nanoscale structural correlations in a model cuprate superconductor}

\author{Zachary W.\ Anderson}
\altaffiliation{These authors contributed equally.}
\affiliation{School of Physics and Astronomy, University of Minnesota, Minneapolis, Minnesota, 55455, USA}
\author{Marin Spai\'{c}}
\altaffiliation{These authors contributed equally.}
\affiliation{Department of Physics, Faculty of Science, University of Zagreb, Zagreb, HR-10000, Croatia}
\author{Nikolaos Biniskos}
\affiliation{School of Physics and Astronomy, University of Minnesota, Minneapolis, Minnesota, 55455, USA}
\author{Liam Thompson}
\affiliation{School of Physics and Astronomy, University of Minnesota, Minneapolis, Minnesota, 55455, USA}
\author{Biqiong Yu}
\affiliation{School of Physics and Astronomy, University of Minnesota, Minneapolis, Minnesota, 55455, USA}
\author{Jack Zwettler}
\affiliation{School of Physics and Astronomy, University of Minnesota, Minneapolis, Minnesota, 55455, USA}
\author{Yaohua Liu}
\affiliation{Second Target Station, Oak Ridge National Laboratory, Oak Ridge, Tennessee 37831, USA}
\author{Feng Ye}
\affiliation{Neutron Scattering Division, Oak Ridge National Laboratory, Oak Ridge, Tennessee 37831, USA}
\author{Garrett E.\ Granroth}
\affiliation{Neutron Scattering Division, Oak Ridge National Laboratory, Oak Ridge, Tennessee 37831, USA}
\author{Matthew Krogstad}
\affiliation{Advanced Photon Source, Argonne National Laboratory, Lemont, IL, 60439, USA}
\author{Raymond Osborn}
\affiliation{Materials Science Division, Argonne National Laboratory, Lemont, IL, 60439, USA}
\author{Damjan Pelc}
\affiliation{School of Physics and Astronomy, University of Minnesota, Minneapolis, Minnesota, 55455, USA}
\affiliation{Department of Physics, Faculty of Science, University of Zagreb, Zagreb, HR-10000, Croatia}
\author{Martin Greven}
\affiliation{School of Physics and Astronomy, University of Minnesota, Minneapolis, Minnesota, 55455, USA}

\date{October 14, 2024}

\begin{abstract}
    Understanding the extent and role of inhomogeneity is a pivotal challenge in the physics of cuprate superconductors.
    While it is known that structural and electronic inhomogeneity is prevalent in the cuprates, it has proven difficult to disentangle compound-specific features from universally relevant effects.
    Here we combine advanced neutron and x-ray diffuse scattering with numerical modeling to obtain insight into bulk structural correlations in HgBa$_2$CuO$_{4+\delta}$.
    This cuprate exhibits a high optimal transition temperature of nearly 100~K, pristine charge-transport behavior, and a simple average crystal structure without long-range structural instabilities, and is therefore uniquely suited for investigations of intrinsic inhomogeneity.
    We uncover diffuse reciprocal-space patterns that correspond to prominent nanoscale correlations of atomic displacements perpendicular to the CuO$_2$ planes.
    The real-space nature of the correlations is revealed through three-dimensional pair distribution function analysis and complementary numerical refinement.
    We find that relative displacements of ionic and CuO$_2$ layers play a crucial role, and that the structural inhomogeneity is not directly caused by the presence of conventional point defects.
    The observed correlations are therefore intrinsic to HgBa$_2$CuO$_{4+\delta}$, and thus likely important for the physics of cuprates more broadly.
    It is possible that the structural correlations are closely related to the unusual superconducting correlations and Mott-localization in these complex oxides.
    As advances in scattering techniques yield increasingly comprehensive data, the experimental and analysis tools developed here for large volumes of diffuse scattering data can be expected to aid future investigations of a wide range of materials.
\end{abstract}

\maketitle

\section{Introduction}

The cuprate high-temperature superconductors rank among the most extensively studied materials systems, yet there remain fundamental open questions, such as the nature of the pairing mechanism.
A pivotal challenge is to unravel the complex relation between structural and electronic properties, in particular given the established presence of disorder in these perovskite-derived compounds.
On the one hand, it has been argued that the essential properties of these materials, including high-$T_c$ superconductivity, can be understood with effective low-energy models that neglect disorder \cite{Abrahams2010}.
On the other hand, various forms of correlated inhomogeneity have been posited to play an essential role for both superconducting pairing and normal-state properties \cite{Egami1994, Phillips2003, Kresin2009, Gomes2007}.
While some of these phenomena are driven by the physics of strongly interacting electrons, the electronic subsystem is coupled to the lattice and structural degrees of freedom are necessarily involved.
Thus, a crucial first step toward a comprehensive understanding of the role of inhomogeneity is detailed structural characterization.
However, it is difficult to experimentally probe nanoscale structural correlations in the bulk, and most cuprate families exhibit idiosyncrasies that lead to non-universal effects and further complicate the picture.

\begin{figure}
    \includegraphics{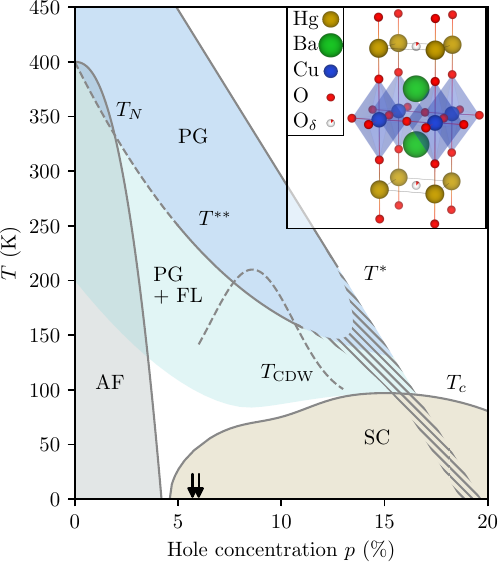}
    \caption{
        Electronic phase diagram of Hg1201 ($p > 0.04$), extended to the undoped insulating state ($p = 0$) based on data for other cuprates \cite{Barisic2013a, Tabis2017}.
        Antiferromagnetic (AF), superconducting (SC), pseudogap (PG) and Fermi-liquid (PG+FL) phases are marked by grey, brown, blue, and cyan shaded areas, respectively.
        Lines indicate the characteristic pseudogap temperatures $T^*$ and $T^{**}$, the superconducting transition temperature $T_c$, and the approximate Neél temperature $T_N$.
        The dashed line at intermediate doping indicates the onset temperature of static charge density wave (CDW) correlations.
        The crystals used for the x-ray and neutron diffuse scattering measurements reported here are underdoped, and have $T_c$ = 50~K and 45~K, respectively, corresponding to $p \approx 6.0\%$ and $5.7\%$ (arrows).
        The sample used for inelastic scattering has $T_c = 71$ K and $p \approx 9.0\%$.
        Inset: schematic crystal structure of Hg1201, with Hg, Ba, Cu, and O atoms colored gold, green, blue, and red, respectively.
        The interstitial oxygen site is indicated by as well and labeled $\mathrm{O}_\delta$.
        \label{fig:hg1201}
    }
\end{figure}

The cuprate phase diagram is controlled by the density of mobile charge carriers in the CuO$_2$ planes, which is typically tuned via chemical substitution or changes in the interstitial oxygen concentration within the intervening rock-salt layers \cite{Eisaki2004}.
The stoichiometric, undoped parent compounds are strongly correlated charge-transfer insulators that become metallic at low carrier concentrations \cite{Ando2004, Ono2007}.
High-$T_c$ superconductivity emerges upon further doping, along with various forms of spin and charge order in the unusual pseudogap state (Fig.~\ref{fig:hg1201}) \cite{Keimer2015} that exhibit varying degrees of real-space inhomogeneity.
Although chemical substitution introduces point disorder, as evidenced, e.g., by bulk nuclear magnetic resonance measurements \cite{Singer2002, Haase2012, Meissner2011, Rybicki2009}, its effects on the metallic CuO$_2$ planes differ widely from material to material.
In addition, there is evidence for different kinds of \emph{intrinsic} inhomogeneity that is not necessarily associated with doping-related point disorder.
For example, short-range charge-density-wave correlations seem to be a generic feature of the phase diagram \cite{Keimer2015, Frano2020}.
Surface-sensitive techniques, such as tunneling microscopy, have enabled detailed characterization of pervasive superconducting gap inhomogeneity that is not directly correlated with point disorder \cite{Alldredge2013, Gomes2007, Du2020, Fischer2007}.
Neutron diffraction measurements indicate that, in some cuprates, the CuO$_2$ planes are buckled at the nanoscale, with the degree of buckling possibly related to superconducting properties \cite{Dabrowski1996, Chmaissem1999}.
Extended x-ray absorption fine structure spectroscopy has shown that local structural distortions affecting the Cu-O bond lengths are present in several cuprates \cite{Bianconi1996, Saini2001}, and it has been suggested that the associated local strains play a crucial role in determining the electronic correlations \cite{Agrestini2003}.
Scale-free structural inhomogeneity and the formation of clusters with distinct local environments have been observed in one cuprate family \cite{Fratini2010}.
Finally, studies of superconducting \cite{Pelc2018, Popcevic2018, Yu2019, Pelc2019} and structural \cite{Pelc2022} fluctuations have uncovered an unusual scaling regime that seems to be caused by underlying, doping- and compound-independent inhomogeneity.
It was recently proposed that such inherent inhomogeneity, in conjunction with a universal transport scattering rate, can explain the unconventional normal-state charge transport and superfluid density behavior \cite{Pelc2019a, Pelc2020}.
Consequently, it is a distinct possibility that intrinsic inhomogeneity plays a pivotal role in cuprate physics and thus highly desirable to obtain direct insight into short-range structural correlations.

Here we use neutron and x-ray diffuse scattering measurements, together with numerical refinement of the local structure, to obtain detailed insight into correlated inhomogeneity in the model cuprate \hbco/ (Hg1201).
We focus on Hg1201 because of its simple crystal structure and representative electronic phenomenology.
In most other cuprates, specific structural characteristics complicate the analysis of local structure and make it difficult to extract underlying universal features.
Examples include structural phase transitions and associated fluctuations in La- and Tl-based cuprates \cite{Egami1994, Axe1994, Wakimoto2006, Pelc2022}, complex superstructure in Bi-based systems \cite{Bianconi1996a}, and oxygen ordering in YBa$_2$Cu$_3$O$_{7-\delta}$ and related compounds \cite{Beyers1989, Veal1990, Jorgensen1990}.
In contrast, Hg1201 exhibits a very simple (tetragonal) average structure and no structural phase transitions.
Similar to other high-$T_c$ cuprates, the structure of Hg1201 involves an alternating stacking of perovskite-based CuO$_2$ planes and ionic rock-salt-like layers, which results in strong anisotropy (Fig.~\ref{fig:hg1201}).
The planes, which consist of a square-planar arrangement of copper-oxygen octahedra with strong hybridization of Cu and O electronic orbitals, are separated by ionic Hg-Ba layers, forming a rather small unit cell with dimensions ($3.8 \times 3.8 \times 9.5$~\AA$^3$).
Hole doping is achieved in Hg1201 via the addition of $\delta$ interstitial oxygen atoms within the ionic layers.
Although there is a tendency for chains of interstitial oxygen atoms to form near optimal doping, this effect is not significantly present at the relatively low doping levels studied in this work ($p \approx 6\%$) \cite{ChabotCouture2010, Izquierdo2011, Welberry2016}.

Furthermore, Hg1201 displays a robust superconducting state with an optimally-doped $T_c$ of 97~K, the highest value among all cuprates with a single CuO$_2$ plane per primitive cell.
Transport measurements indicate minimal disorder effects (e.g., due to oxygen interstitials) on the electronic physics in Hg1201: the low-temperature residual resistivity is tiny \cite{Chan2014, Barisic2019}; quantum oscillations associated with a single electron pocket are seen in high magnetic fields in the underdoped part of the phase diagram \cite{Barisic2013, Chan2016b}; in-plane magnetoresistance measurements reveal simple Kohler scaling in the normal state \cite{Chan2014}; and stripe/spin-density wave correlations are absent in the magnetic response \cite{Chan2016, Chan2016a}.
While charge-density-wave (CDW) order is seen at moderate doping levels, as in other cuprates, much of the present work focuses on samples at lower doping, where CDW order is strongly suppressed \cite{Tabis2014, Tabis2017} and the interstitial oxygen density is small.
Hg1201 is therefore an excellent model system for studies of intrinsic nanoscale structural correlations.

In order to achieve detailed understanding of the local structure of Hg1201, we take full advantage of recent advances in diffuse scattering techniques and instrumentation at both neutron and synchrotron x-ray facilities.
We use thermal neutron and hard x-ray diffraction to probe structural correlations in the bulk of single-crystal samples.
In diffraction experiments, the Bragg peaks, which form the reciprocal lattice, encode information about the average (i.e., long-range) structure of a crystal, whereas local deviations from the average structure result in scattering away from the reciprocal lattice vectors.
This diffuse scattering can be quasielastic (i.e., due to quasi-static distortions) or inelastic (due to excitations), and it can result from structural, electronic, or magnetic short-range correlations.
In a system with short-range correlations, diffuse scattering will show distinct reciprocal-space patterns, yet it is highly nontrivial to determine the real-space structure from experiment, as it is necessary to gather extremely high-quality data for many Brillouin zones.
We use two complementary approaches to resolve the real-space correlations: the vector pair distribution function (\tdpdf/) transform and reverse Monte Carlo refinement.
The \tdpdf/ transform is a nearly assumption-free method to visualize real-space correlations between atomic pairs \cite{Weber2012} and has been used in recent studies of correlated inhomogeneity in ionic conductors, thermoelectric materials, bismuthate superconductors, etc. \cite{Sangiorgio2018, Krogstad2020, Griffitt2023}.
Here we generate \tdpdf/s from both neutron and x-ray data, which enables detailed insight due to the differing atomic form factors of the two probes.
In addition, we establish a computational framework to perform large-volume reverse Monte Carlo simulations of the real-space structure, with fits to both neutron and x-ray reciprocal-space data.
This method reveals correlations that are hidden in the \tdpdf/ due to atomic pair multiplicity and overlap, and our combined measurement and simulation approach enables us to reliably resolve the salient local-structure features in Hg1201.
In particular, we uncover a pattern of correlated atomic displacements that predominantly involve the ionic layers and apical oxygen atoms of the Cu-O octahedra.
Given that these nanoscale correlations are associated with generic features of the cuprate structure, they are likely independent of the structural details of the system, with potentially far-reaching implications for the physics of the cuprates.
We expect that the methods developed here will be useful for advanced quantitative analysis of local structure in a broad range of materials systems.

This manuscript is organized as follows: In Section II we describe various experimental details, in Section III we present our results and analysis (reciprocal-space features, \tdpdf/, modeling), and in Section IV we discuss and summarize our findings.

\section{Experimental details}

Single crystals of Hg1201 were grown with the self-flux method described in detail in \cite{Zhao2006}.
Briefly, HgO powder and a crucible containing BaCuO$_3$ powder were sealed under vacuum in a quartz tube and heated in a furnace first to 800$^\circ$C, then to 1020$^\circ$C, followed by a slow cooling step to produce large crystals.
This method typically produces single crystals with masses ranging from 5 - 100 mg.
The sample used for the diffuse neutron scattering experiment was a single crystal with visible $ab$ plane facets and a mass of 90 mg, while that used for the diffuse x-ray scattering measurements had a mass of $\sim$5 mg.
In order to control the interstitial oxygen concentration, and thereby the hole concentration, the crystals were annealed for four weeks at a temperature of 475$^\circ$C in vacuum, produced using a rotary vane pump with base pressure of 100 mTorr.
The large single crystal used for diffuse neutron scattering was covered with as-grown Hg1201 powder during the anneal, while the smaller crystal used for x-ray scattering were annealed uncovered. This resulted in respective transition temperatures of $T_c = 45$~K ($p \approx 5.7\%$) and $T_c = 50$~K ($p \approx 6.0\%$).
As noted, these doping levels are below the range in which CDW order has been observed in Hg1201 \cite{Tabis2017}.
Inelastic neutron scattering was used to measure the excitation spectrum in regions of reciprocal space near significant diffuse scattering intensity.
In this case, the sample was a co-mounted array of ten moderately underdoped ($T_c = 71$~K, $p \approx 9.0\%$) crystals with a total mass of 0.5 g.
These crystals were annealed uncovered for 80 days at 400$^\circ$C, using the same hardware and vacuum atmosphere.
$T_c$ was measured via DC magnetization in a Quantum Design, Inc.
Magnetic Properties Measurement System.
To estimate hole concentration from the measured $T_c$, we used a linear interpolation of published results for $T_c(p)$ based on electronic and thermoelectric transport measurements \cite{Yamamoto2018}.

X-ray diffuse scattering measurements were performed on beam line 6-ID-D at the Advanced Photon Source, Argonne National Laboratory.
We used hard x-rays with an energy of 87 keV, generated by a superconducting undulator, and the scattering was measured in transmission geometry with a Pilatus 2M CdTe detector that has a high dynamic range for high-energy photons.
The use of a high photon energy has two distinct advantages: relatively low absorption, even for mm-sized crystals, and access to thousands of Brillouin zones, even in systems with small primitive cells like Hg1201.
The last feature is essential to obtain a reliable reconstruction of real-space correlations.
We employed an experimental setup and procedure similar to that in \cite{Krogstad2020}.
In order to maximize the relatively weak diffuse scattering signal, the incident beam was not attenuated during the measurements, which resulted in inaccurate intensities of strong Bragg peaks that saturated the detector.
In addition, some strong Bragg peaks produced artifacts in the detector from scattering within the CdTe sensor layer; these artifacts were masked prior to further analysis using a procedure described in the Supplementary Information of Ref.~\cite{Krogstad2020}.
Three data sets were collected while rotating the sample 360$^\circ$, with the rotation axis at angles 75$^\circ$, 90$^\circ$, and 105$^\circ$ relative to the incident beam, and then combined to improve the counting statistics and fill gaps that resulted from the aforementioned masking procedure.
The data were then symmetrized according to the sample's average-structure symmetry, $P4/mmm$, which resulted in a total data set spanning a large reciprocal space volume of $\sim$10$^4$ Brillouin zones centered on the origin, $\mathbf{Q} = (0~0~0)$.
Due to the large coverage of reciprocal space, the initial sample orientation did not have a significant effect on the final analysis, and therefore the sample was not mounted in a specific high-symmetry orientation.
Temperature control was provided by a cold helium stream blown over the sample, and the experiment was performed at the base temperature of 30 K in order to minimize thermal diffuse scattering.

Neutron diffuse scattering measurements were performed on beam line BL-9 (CORELLI) at the Spallation Neutron Source, Oak Ridge National Laboratory.
This dedicated spectrometer employs a specialized cross-correlation technique to discriminate between quasielastic and total scattering, while using a broad-spectrum thermal neutron beam \cite{Ye2018}.
The incident neutron beam is thus used very efficiently compared to monochromation-based techniques, while still allowing for the quasielastic scattering to be separated from the total scattering.
This results in an energy resolution of roughly 1 meV centered at the elastic line.
Temperature control was provided by a closed-cycle refrigerator, and the experiment was performed at the base temperature of 5 K, again to minimize thermal diffuse scattering.
The sample was mounted on an aluminum post using cyanoacrylate glue and aluminum foil wrapping, and cadmium foil was employed to mask the glued section of the crystal to avoid illuminating the glue with the neutron beam.
The high-symmetry planes of the crystal were offset from the scattering plane in order to provide maximum reciprocal space coverage after a symmetrization of the data according to the space group of the average structure ($P4/mmm$).

Inelastic neutron scattering was performed on BL-18 (ARCS) at the Spallation Neutron Source, Oak Ridge National Laboratory \cite{Abernathy2012}.
The crystals were co-mounted on an aluminum sample holder using GE varnish and measurements were performed with the sample heated to a somewhat elevated temperature of 345~K in order to enhance the inelastic response.
The Fermi chopper slit package has a 1.5~mm slit width and was spun at 180~Hz. It was phased for 70~meV neutrons.
The sample was mounted in the $[1~0~0]/[0~0~1]$ scattering plane and rotated over 45$^\circ$ during the measurement in order to measure a region of reciprocal space near $(0~0~10)$.
The data were reduced using the Mantid framework \cite{Mantid2014}.
Variations of sensitivity in the ARCS detector bank were corrected for by using a white beam measurement of a vanadium standard \cite{ARCSnorm2021a}.

\section{Results and analysis}

\subsection{Reciprocal-space}

\begin{figure*}
    \includegraphics{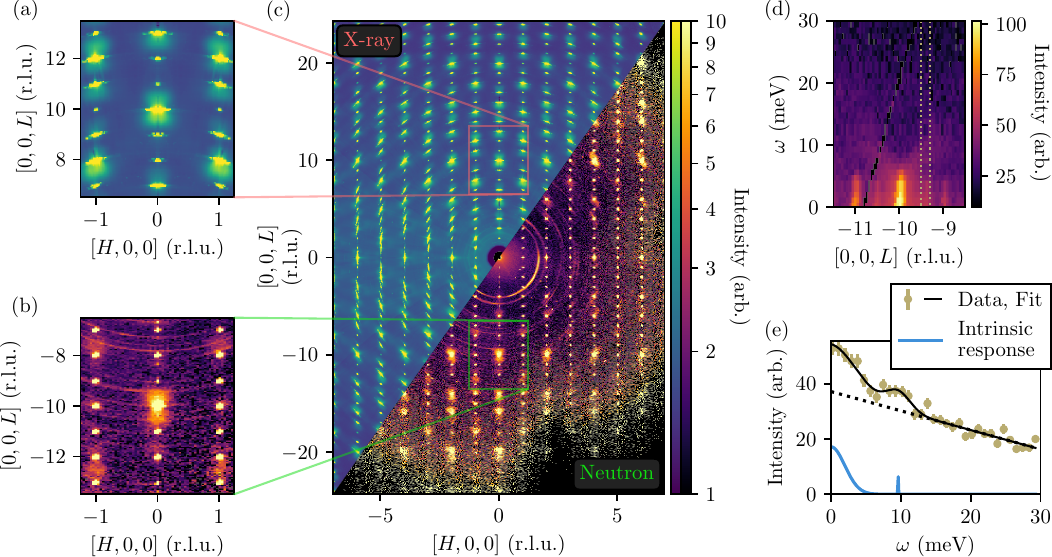}
    \caption{
    Representative scattering data.
    (a) X-ray scattering results centered on the (0 0 10) Bragg peak, showing diffuse scattering primarily as lobes of intensity extending along [0 0 1] from certain Bragg peaks.
    Data collected at 30~K for a $T_c = 50$~K sample.
    (b) Quasielastic neutron scattering results centered on the (0 0 -10) Bragg peak. Similar diffuse scattering is seen near the (0 0 10) peak.
    Data collected at 5~K for a $T_c = 45$~K sample.
    (c) Scattering intensity in the $[H 0 0]/[0 0 L]$ plane, showing x-ray (left/top) and neutron (right/bottom) scattering results.
    On the arbitrary units scale used in (a-c), the x-ray Bragg peak intensities are typically $\sim 600$, and the neutron Bragg peak intensities are typically $\sim 1000$.
    (d) Slice of inelastic neutron scattering data with $\mathbf{Q} \parallel [0 0 L]$ at $T = 345$~K for a $T_c = 71$~K sample, showing the distinct quasi-elastic diffuse scattering and inelastic phonon dispersion.
    Dotted lines indicate the momentum range used in (e).
    (e) Energy dependence of the scattered intensity in the narrow $\mathbf{Q} = (0~0~L)$ range with $L = -9.4 \pm 0.1$ r.l.u. as indicated in (d).
    The solid black line shows a fit to two gaussian peaks, one centered at $\omega = 0$ and the other at $\omega \approx 10$ meV,  and a linear background (dotted line).
    The solid blue line indicates the fit result after removing the linear background and accounting for the broadening effects of instrumental resolution calculated using McVINE \cite{McVINE1, McVINE2}.
    \label{fig:qspace_arcs}
    }
\end{figure*}

As seen from Fig.~\ref{fig:qspace_arcs}(a-c), the main feature in both the neutron and x-ray diffuse scattering data is a set of broad, bow-tie-shaped lobes of intensity that extend along $[001]$ from many Bragg peaks.
The detailed structure and large-$Q$ dependence differs between the two datasets; the diffuse scattering is generally stronger at large $Q$, indicative of a likely origin in atomic displacements, rather than correlated occupational disorder.
The bow-tie-like features are reminiscent of Huang scattering, which has been attributed to the diffuse scattering from intermediate-range strain fields near point defects, e.g., in doped semiconductors \cite{Huang1947, Krivoglaz1969, Barabash1999}.
In contrast to conventional Huang scattering, we observe a strong asymmetry in the high-$Q$ vs. low-$Q$ intensities for some of the lobes.
This indicates that local lattice expansion and/or contraction must take place, as discussed in more detail below.

\subsubsection{Contrast between x-ray and neutron scattering}

The differences in the x-ray and neutron diffuse scattering patterns reflect the distinct energy resolution and scattering interactions with the system under study.
In particular, CORELLI allows the efficient use of the incident neutron beam with an energy resolution of either $\sim 1$ meV (quasielastic scattering) or $\sim 10$ meV (total scattering), whereas the x-ray experiment effectively integrates over all energy scales (tens of eV), thereby probing the instantaneous correlations of electrons, including dynamic correlations such as those related to phonons.
The two probes also feature important differences in structure factors and atomic form factors.
Whereas x-rays scatter primarily from charge density, and are more sensitive to heavier elements (Ba and Hg in the case of Hg1201), neutrons are more sensitive to light elements (O in the case of Hg1201).
Despite these differences, the x-ray and neutron data in Fig.~\ref{fig:qspace_arcs} have common features, and the reciprocal-space structures are remarkably similar.
This indicates that the diffuse scattering results from intermediate-range, quasi-static correlations, and as such is captured with similar fidelity by probing the electron or nuclear distributions.

\subsubsection{Dynamic versus quasielastic response}

Comparison of the neutron quasielastic ($\sim$1 meV resolution) and total scattering (energy-integrated, $\sim$10 meV resolution) datasets provides basic, qualitative insight into the energy dependence of the response:
We find that the quasielastic diffuse scattering is approximately coincident with reciprocal space regions of more intense inelastic diffuse scattering.
In order to further investigate this, we used the ARCS neutron spectrometer at Oak Ridge National Laboratory to discern the excitation spectrum in several Brillouin zones.
This measurement indicates that the inelastic diffuse scattering is primarily caused by low-energy acoustic phonon modes that are clearly separated in energy from the quasielastic diffuse scattering, as indicated in Fig.~\ref{fig:qspace_arcs} (d, e).
When the instrument resolution is taken into account, the low-energy acoustic phonon modes are found to be resolution-limited and distinct from the quasielastic diffuse scattering.

\subsection{3D-\texorpdfstring{$\Delta$}{Δ}PDF results}

The \tdpdf/ technique involves the separation of diffuse scattering and Bragg scattering, and the subsequent Fourier transformation of the former yields a pair distribution function of correlations in the deviations from the average real-space structure \cite{Weber2012}.
The Bragg and diffuse scattering can only be separated approximately, and thus the \tdpdf/ technique is difficult to use for a quantitative analysis of local structure, although it is a useful tool to reveal correlations without any modeling of the local structure.
In the present work, we use the ``punch-and-fill'' method, whereby spherical regions centered at Bragg peaks are ``punched'' out of the data set and then filled in by convolving the remaining data with a gaussian kernel.

The real-space resolution of the \tdpdf/ is governed both by the extent of the data in reciprocal-space, i.e., $\mathrm{max}(Q)$, and the form of the taper function used to reduce leakage artifacts in the Fourier transform.
There is also an intrinsic contribution from the thermal broadening produced by atomic vibrations.
This makes it difficult to quantitatively determine the resolution and sensitivity of the \tdpdf/ from first principles.
In practice, the spatial resolution of the \tdpdf/ can be approximated by considering the width of its sharpest features.
By this reasoning, the effective resolution of the \tdpdf/ for the x-ray-based results is $\sim$0.3 \AA, compared to $\sim$0.4 \AA\, for the neutron-based results.

\begin{figure*}
    \includegraphics{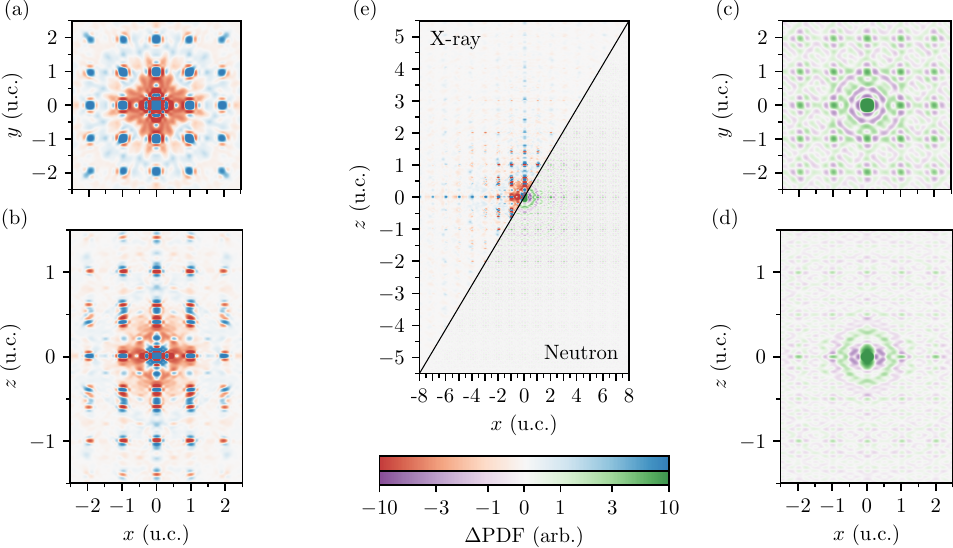}
    \caption{
        \tdpdf/ calculated from the x-ray and neutron diffuse scattering data shown in Fig.~\ref{fig:qspace_arcs}(a-c).
        (a,b) The $xy0$ and $x0z$ planes of the \tdpdf/ calculated from x-ray scattering data, zoomed in to show the region near (0 0 0).
        (c,d) The corresponding $xy0$ and $x0z$ planes of the \tdpdf/ calculated from neutron scattering data.
        (e) A zoomed-out view of the $(x0z)$ plane of the \tdpdf/, showing both x-ray (top/left) and neutron (bottom/right) based results.
        \label{fig:pdf}
        The top (red-blue) and bottom (purple-green) color bars indicate the scales used for x-ray- and neutron-based \tdpdf/ results, respectively.
        The same color scales are used in all panels.
    }
\end{figure*}

The \tdpdf/ results clearly reveal several general features of the local correlations.
Figure \ref{fig:pdf} shows 2D planar cuts along high-symmetry directions of the \tdpdf/s calculated from both x-ray and neutron scattering data.
The coordinates $x$ and $y$ parameterize correlations along directions parallel to the Cu-O bonds in the CuO$_2$ plane, and the $z$ coordinate parameterizes out-of-plane correlations (see Fig.~\ref{fig:hg1201} inset).
The sign and value of the \tdpdf/ indicates the extent to which correlations are stronger or weaker than in the average structure.
We find relatively little correlated distortions parallel to the CuO$_2$ planes, as in this case the \tdpdf/ is dominated by contributions characteristic of thermal diffuse scattering (i.e., a positive peak at integer $(x,y,z)$ surrounded by a negative shell).
On the other hand, local atomic shifts along the $c$-axis are more prominent, characterized by structures with negative PDF near the average-structure positions with positive PDF nearby.
These $c$-axis atomic displacements have distinct correlation lengths in the $ab$ and $ac$ planes.
While the full detail of the \tdpdf/ can not be accurately captured by a simple anisotropic correlation length (e.g., with ellipsoidal dependence of correlations on real space direction), the typical extent of local correlations can be described approximately with correlation lengths of $\sim 7$ unit cells parallel to the CuO$_2$ planes and $\sim 3$ unit cells along the $c$-direction.
Stronger correlations are present along the $a$, $b$, and $c$ directions (i.e., when $x = 0$, $y = 0$, or $z = 0$), though the length scale over which these correlations persist is approximately the same.
It is also clear from some features near half-integer positions along $[0~0~1]$ that the diffuse scattering results in part from shifts of Ba atoms along the $c$-axis, as evidenced by adjacent positive and negative peaks in the \tdpdf/ near the Ba-Ba pair vectors, e.g., (0 0 0.4) and (0 0 0.6).
Due to the overlap of multiple atomic pair vectors at most other common positions (e.g., integer or half-integer $x$, $y$, and $z$), it is difficult to unambiguously determine the local distortions of other atomic pairs.

\subsection{Modeling}

\begin{figure*}
    \includegraphics{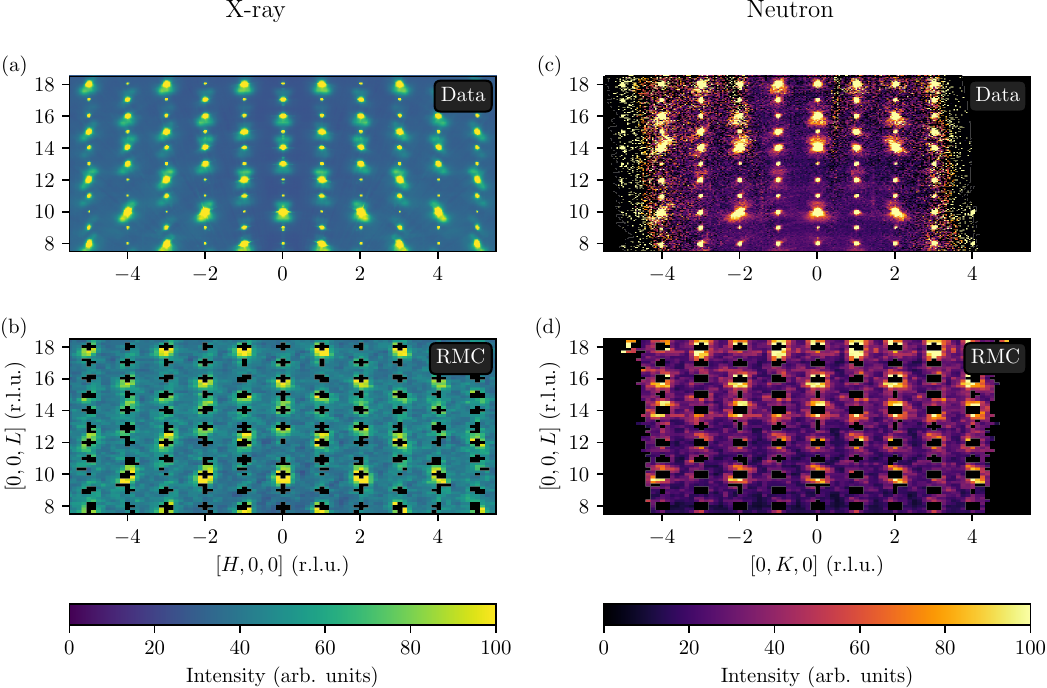}
    \caption{
        Comparison between data (top) and simulated diffuse scattering (bottom) from a structure refined to the same data using the reverse Monte Carlo procedure described in the text, for both (a,b) x-ray and (c,d) neutron data.
        The data panels show the same data sets as in Fig.~\ref{fig:qspace_arcs}.
        The panels show a section of the $[H~0~L]$ plane (x-ray) or $[0~K~L]$ plane (neutron) integrated over the range $-0.04~\mathrm{r.l.u.} \le q \le 0.04~\mathrm{r.l.u}$ in the direction normal to the plane, i.e., along $K$ and $H$ for x-ray- and neutron-based panels respectively.
        The black regions in (b, d) indicate where were data removed prior to RMC refinement via a statistical outlier method so as to refine the local rather than average structure.
        \label{fig:compare_rmc}
    }
\end{figure*}

A shortcoming of the \tdpdf/ method is its inability to reliably resolve the degrees of freedom responsible for diffuse scattering in situations where there is a significant overlap of pair separation vectors that correspond to different atomic pairs in the structure.
Since this is the case for Hg1201, we resorted to numerical modeling to complement the insights obtained via the \tdpdf/ method. In particular, we used a variant of the reverse Monte Carlo (RMC) method \cite{McGreevy1988, McGreevy2001} in order to systematically generate real-space models of Hg1201 local structure by stochastically refining them against the relevant portions of experimental diffuse scattering data \cite{Proffen1997}.
In order to capture both the details of diffuse features around individual Bragg positions and their intricate distribution over different zones, we refined our models against a sufficiently large, four-fold symmetric section of reciprocal space ranging from -5.5 to 5.5 reciprocal lattice units (r.l.u.) along symmetry-equivalent $[1~0~0]$ and $[0~1~0]$ directions, and from $7.5$ to $18.5$ r.l.u. along $[0~0~1]$.
Both x-ray and neutron intensities were rebinned to a lower resolution of $8 \times 8 \times 6$ bins per zone prior to refinement, where the binning resolution was chosen to be sufficiently large to preserve information about the relevant intermediate range ($\sim$ 3-4 unit cells) correlations visible in the \tdpdf/ (Fig.~\ref{fig:pdf}) and small enough to make the RMC refinement computationally viable.
Furthermore, all Bragg-peak intensities were excluded (``punched out'') from the data using a simple statistical outlier elimination method outlined in \cite{Morgan2021} to avoid fitting the partly saturated experimental Bragg intensities and to focus the refinement on the local structure.
In order to improve the statistical significance of our results, 30 concurrent refinements were performed against both neutron and x-ray data with independent model supercells, each of size $8 \times 8 \times 6$ unit cells in order to match supercell Bragg positions to the reciprocal space grid of the rebinned data.
It is worth noting that such a compromise \textit{a priori} limits our ability to reliably resolve correlations on scales larger than 3-4 unit cells.
Nevertheless, these longer-range correlations likely correspond to distortion fields induced by correlations on shorter length scales and therefore do not contain fundamentally new physical effects.
Also, since no direct evidence of correlations induced by oxygen interstitials is seen in the \tdpdf/, we chose to refine \hbco/ model supercells with $\delta = 0$, thereby focusing exclusively on displacive disorder and leaving open the question of possible indirect influence of occupational disorder.
Additionally, this assumption is justified \textit{a posteriori} by the satisfactory match of refined and model intensities (see Fig.~\ref{fig:compare_rmc}).
During refinement, disorder is introduced in each step by shifting a randomly selected atom away from the average position by a random displacement vector drawn from a multivariate gaussian distribution with the covariance matrix based on the measured \cite{Pissas1997} atomic displacement parameters of Hg1201.
This was done to ensure that the proposed RMC moves automatically preserve the one-point correlations associated with the known thermal ellipsoids, while allowing two-point correlations responsible for diffuse scattering to be optimized by the refinement.
Diffuse scattering intensity $I_{\text{c}}(\mathbf{q})$ was then calculated on the reciprocal space grid chosen for refinement with periodic boundary conditions imposed to minimize the finite-size effects and convolved with a narrow gaussian kernel to eliminate the high-frequency Fourier noise inherent in all numerical diffuse scattering calculations \cite{Welberry1992, Paddison2019}.
This step represents the main computational bottleneck of the refinement and was made viable in our implementation by the use of state-of-the-art non-uniform fast Fourier transform routines in the \texttt{FINUFFT} software library \cite{FINUFFT} and its GPU-accelerated counterpart \cite{cuFINUFFT}.
This development enabled the efficient ($\sim \mathcal{O}(n \log n)$) recalculation of diffuse scattering in each RMC step and allowed us to circumvent the small, but hard-to-control numerical drift inherent in recently proposed \cite{Morgan2021}
RMC implementations based on updating the scattering amplitude after each RMC step.
Monte Carlo importance sampling was implemented by accepting the proposed displacement with probability $P$ given by
\begin{equation}
    \label{p}
    P =
    \begin{cases}
        1                    & \text{if } \Delta\chi^2 \le 0 \\
        e^{-\Delta \chi^2/T} & \text{if } \Delta\chi^2 > 0
    \end{cases}
\end{equation}
where $T$ is the effective Monte Carlo temperature used to control the acceptance ratio, and the fit parameter $\chi^2$ is defined as
\begin{equation}
    \label{chi}
    \chi^2
    =
    \sum_{\mathbf{q}}\left[
        \frac
        {(a I_{\text {c}}(\mathbf{q}) + b) - I_{\text {e}}(\mathbf{q})}
        {\sigma_{\text {e}}(\mathbf{q})}
        \right]^2 \,.
\end{equation}
Here, $I_{\text {e}}(\mathbf{q})$ is the rebinned experimental scattering intensity and $\sigma_{\operatorname{e}}(\mathbf{q})$ are the experimental uncertainties associated with $I_{\text {e}}(\mathbf{q})$ which are used to weigh the importance of individual data points.
The constants $a$ and $b$ in Eq.~\eqref{chi} are used to account for the overall scale and flat background mismatch between $I_{\text {c}}(\mathbf{q})$ and $I_{\text {e}}(\mathbf{q})$ and calculated using least-squares linear regression formulas before calculating $\chi^2$.
The procedure described above was repeated $N_a$ times per refinement cycle, where $N_a$ denotes the total number of atoms in a given model Hg1201 supercell.
Additionally, a simulated annealing metaheuristic \cite{Press2007} was implemented by slowly reducing the effective temperature in Eq.~\eqref{p} during the refinement according to the exponential cooling schedule defined recursively as
\begin{equation}
    T_{i+1} = T_i e^{-\lambda}\,,
\end{equation}
where $T_i$ is the effective temperature in the $i$-th RMC step and $\lambda$ is the cooling rate.
In our case, 400 RMC refinement cycles were performed for each model supercell, with the empirically adjusted cooling rate $\lambda = 10^{-4}$.
In this way, the acceptance ratio and, consequently, the explored configuration space were reduced slowly enough to prevent simulated atomic positions from being pinned in one of the local minima of $\chi^2$ as the refinement converges.

Having obtained a satisfactory description of the data (see Fig.~\ref{fig:compare_rmc}), we turn to investigate the short-range correlations that comprise the local structure of Hg1201.
We start by calculating the normalized displacement correlation coefficient
\begin{equation}
    \label{corr}
    \left\langle A^\alpha B^\beta \right\rangle
    =
    \frac
    {\left\langle {\Delta x}_{i\alpha} {\Delta x}_{j\beta}\right\rangle}
    {\sqrt{
            \left\langle{\Delta x}_{i\alpha}^2\right\rangle
            \left\langle{\Delta x}_{j\beta }^2\right\rangle
        }}
    \,,
\end{equation}
for different atomic pairs ($AB$) and atomic displacement components (${\Delta x}_{\alpha}$, ${\Delta x}_{\beta}$) such that the indicated averages are performed both over equivalent atomic pairs in a single supercell and over all refined supercells.
For example, $\left\langle \mathrm{Hg}^z \mathrm{O}_\mathrm{a}^x \right\rangle$ denotes the normalized correlation between the displacement of Hg atoms along the $z$-axis and apical oxygen atoms along the $x$-axis.

\begin{figure*}
    \includegraphics{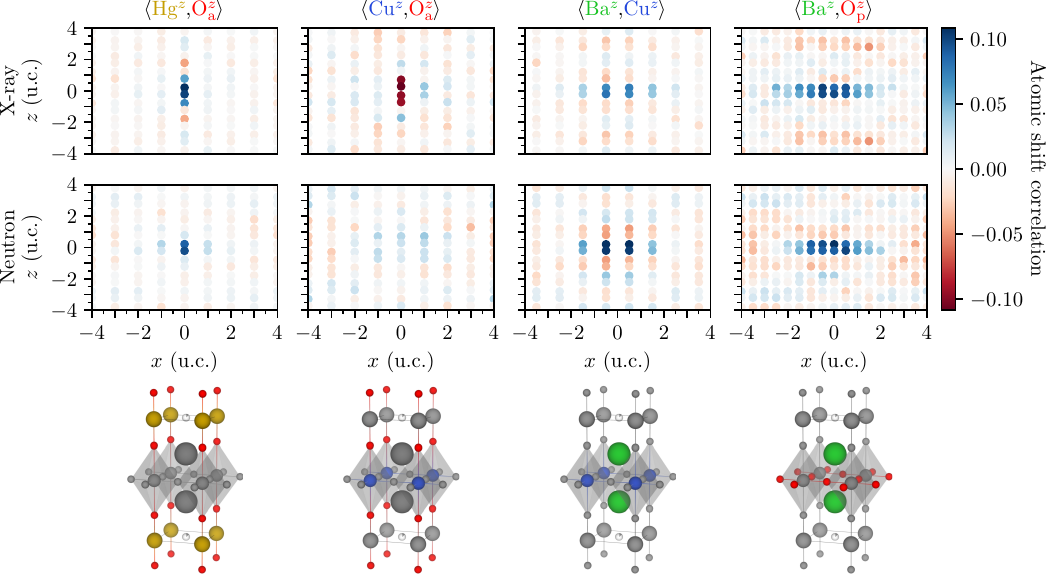}
    \caption{
        Maps of the correlations of atomic shifts calculated from the RMC-refined structures.
        Each column shows $\left\langle A^z B^z \right\rangle$, the correlations between the shifts of the two indicated atoms $A$ and $B$ along the $z$ axis (see Eq.~\eqref{corr} and main text for details).
        The upper panels show results from x-ray-based refinements, and the middle panels show results from neutron-based refinements.
        Each dot indicates an average-structure atomic pair vector, and the color of the dot shows the sign and strength of the correlation between local displacements of the atoms in that pair along the $z$ direction.
        Structure diagrams in the bottom row indicate the atomic sites under consideration, highlighted in color.
        \label{fig:rmc_correlations_main}
    }
\end{figure*}

As expected from the reciprocal-space distribution of the diffuse scattering and from the \tdpdf/ analysis, significant (positive and negative) correlations appear overwhelmingly between out-of-plane ($z$-direction) displacements, which indicates that these play a dominant role in the local structure.
We find that the strongest of these are positive out-of-plane ($zz$) correlations between nearest-neighbor Hg and apical oxygen (O$_\mathrm{a}$) atoms.
RMC refinements based on the x-ray scattering data also show significant negative correlations for Cu-O$_\mathrm{a}$ pairs along $(00z)$, although we do not see such correlations in the neutron-based RMC refinements.
This discrepancy may be a result of the smaller reciprocal-space extent and lower signal-to-background ratio of the neutron scattering data, which lowers the sensitivity of the neutron-based RMC refinements to small correlated distortions.
Together, these correlations suggest that the local structure of Hg1201 results in large part from an interplay among coordinated $z$-direction shifts involving Hg-O$_\mathrm{a}$ pairs or O$_\mathrm{a}$-Hg-O$_\mathrm{a}$ triples and possible out-of-plane distortions of Cu-O octahedra involving opposite shifts of Cu and O$_\mathrm{a}$.
Ba-related pair correlations also turn out to be significant, notably shifts of Ba along the $z$-axis relative to Cu and planar oxygen (O$_\mathrm{p}$) atoms, whose strongest correlations are predominantly parallel to the CuO$_2$ planes.
The correlations involving the aforementioned pairs are shown as a function of average-structure pair separation vector in the $(x0z)$ plane in Fig.~\ref{fig:rmc_correlations_main}.

\section{Discussion and summary} \label{discussion}

Our diffuse scattering study reveals a remarkable number of structured reciprocal space features, despite the simple average structure of Hg1201.
The most important question then concerns the universality of these features: are they specific to this compound, or of wider relevance to the cuprates?
While only further measurements of other cuprates can conclusively answer this question, we can make several inferences from the Hg1201 data.
First, we do not observe significant direct or indirect signatures of correlated displacements that would be caused by occupational disorder associated with mercury vacancies or interstitial oxygen atoms.
If either type of occupational disorder were spatially correlated, it would produce structured diffuse scattering with significant intensity at low Q, which is not observed.
Similarly, if the structured diffuse scattering was the result of mercury vacancies or oxygen interstitials, it should be visible as local in-plane correlations among the Hg atoms.
However, such correlations are much weaker than the predominant out-of-plane displacements, which indicates that the latter are intrinsic and independent of occupational disorder.
It is, of course, possible that the presence of occupational disorder influences the observed structural correlations in more subtle ways: e.g., the interstitial-induced random fields could serve as nucleation sites or pinning potentials for intrinsic correlations, and their spatial distribution could affect the correlation length.
Indeed, the $ab$-plane correlation length estimated from diffuse scattering ($\sim 7$ unit cells) is similar to the typical in-plane distance between interstitials ($\sim$ 5 - 10 unit cells).
Second, the apical oxygen atoms seem to play an important role in the local structural distortions.
There are at least two distinct possibilities for the microscopic origin of this behavior: it could be due to Hg-O interactions, i.e., the properties of the ionic layer, or due to intrinsic distortions of the Cu-O octahedra.
There is no reason for either mechanism to be specific to Hg1201, given the near-universal presence of ionic layers and Cu-O octahedra (or similar structures) in superconducting cuprates.

It is known that the ionic layers of the cuprates are highly polarizable \cite{Mallett2013, Reagor1989}, and the Hg-O scenario corresponds to the interesting possibility of local electric dipole formation.
If a dipole appears in a given ionic layer, the conducting Cu-O planes would screen it and induce dipoles of opposite sign in the neighboring ionic layers, leading to an antiferroelectric arrangement along the crystallographic $c$-axis.
The correlation plots in Fig.~\ref{fig:rmc_correlations_main} show evidence for just such an effect: the sign of the Hg-O correlations changes from nearest-neighbor to next-nearest-neighbor pairs.
Short-range antiferroelectric order could play a significant role in cuprate physics since it would create nontrivial inter-plane interactions.
It has been suggested that the ionic layer could be crucial for cuprate superconductivity \cite{Mallett2013}, hydrostatic pressure experiments on Hg1201 and La$_{2-x}$Sr$_x$CuO$_4$ have shown that superconducting features are not uniquely determined by the properties of the Cu-O planes alone \cite{Wang2014}, and there are signs of macroscopic electric polarization hysteresis in some strongly underdoped cuprates \cite{Viskadourakis2012}.
Comparative studies of several cuprate families with different ionic layers will be important to conclusively establish if local polarization is indeed the underlying cause of the correlated inhomogeneity.

The other possible mechanism for the observed displacements involves out-of-plane distortions of the Cu-O octahedra, which would then induce the Hg-O correlations.
Similar effects have been found in several transition-metal oxides with octahedral coordination, such as manganites, where they are associated with the formation of polarons \cite{Dai2000, Campbell2003}.
In that case, charges self-localize due to strong electron-phonon interactions, which leads to local octahedral distortions.
The coupling of electrons to some optical phonon branches is relatively strong in the cuprates \cite{Vishik2014, McQueeney1999}, and polaronic physics has long been suggested to be relevant \cite{Salje1995, Bishop2003, Weyeneth2017}.
While it is unlikely that the mobile charge carriers are of polaronic character \cite{Chakraverty1998, Chan2014, Barisic2019, Pelc2020}, the presence of additional localized charges could well be associated with lattice distortions.
Indeed, magnetic resonance \cite{Meissner2011, Haase2012, Haase2022}, specific heat \cite{Storey2012}, and transport \cite{Ayres2021} experiments have demonstrated the coexistence of two electronic subsystems - a Fermi-liquid-like metal, and an unconventional strongly-correlated component \cite{Ayres2021}.
The latter has been related to Mott physics and localization and seems to play a crucial role in the superconducting pairing mechanism \cite{Pelc2019a, Pelc2020, Ayres2021, Aavishkar2023}.
It is an intriguing possibility that the observed structural correlations are associated with the localized component, as this would imply a fundamental role of inhomogeneity in this key aspect of cuprate physics.
Further investigation is needed to firmly resolve the origin of the correlations and to directly test their relation to the electronic subsystem.
We expect that studies of the response to lattice deformation induced externally by hydrostatic pressure or uniaxial strain, with complementary electronic measurements will be especially illuminating.

Importantly, we find no evidence of CuO$_2$ plane buckling in our data, which would appear as negative correlations between shifts of Cu and in-plane O atoms.
Neutron scattering would be especially sensitive to such effects due to the relatively large oxygen scattering length.
In fact, we find quite strong \emph{positive} correlations within the planes, implying that they respond as fairly rigid flat sheets.
This observation contrasts with previous neutron powder diffraction studies, where indications of buckling were obtained in several cuprate families, with a proposed relation between $T_c$ and local buckling angle \cite{Dabrowski1996, Chmaissem1999}.
This relation between $T_c$ and buckling angle was observed in two families of cuprates which are prone to orthorhombic structural distortions, and it may be a peculiarity of compounds with these structural complications, rather than a fundamental property of the cuprates more generally.
Our single crystal data enable detailed three-dimensional insight, and leave little doubt that, at least in highly-underdoped samples of simple-tetragonal Hg1201, correlated out-of-plane distortions are more prominent than correlated in-plane distortions associated with buckling.

Recent work has uncovered doping- and compound-independent features in the superconducting fluctuation regime \cite{Pelc2018, Popcevic2018, Yu2019, Pelc2019}, with similar behavior also found for fluctuations associated with the structural transition in La- and Tl-based cuprates that involves a staggered tilting of the CuO$_6$ octahedra \cite{Pelc2022}.
The observed unusual exponential temperature dependences were related to the formation and proliferation of locally ordered regions, precipitated by the presence of underlying inhomogeneity.
In the present work, we observe correlated inhomogeneity \emph{directly}, and the short-range correlations we uncover in Hg1201 could plausibly be responsible for the appearance of the universal unconventional electronic fluctuation regime.
The local distortions could therefore substantially affect superconductivity, especially through a modulation of the inter-plane coupling, as the local structure has a much longer $c$-axis correlation length than the corresponding superconducting coherence length, and these length scales are similar for correlations/coherence in the $ab$ plane \cite{Hofer1998}.
In contrast, we do not expect significant effects on the in-plane normal-state mobility, given the small distortions of the planar Cu-O bonds and the absence of buckling.
This is in line with the very low residual resistivity values observed in high-quality Hg1201 crystals \cite{Chan2014,Barisic2013,Chan2016b}.

In summary, we have uncovered significant correlated nanoscale structural inhomogeneity in simple-tetragonal \hbco/ associated with relative out-of-plane shifts of the Hg, apical O, and Cu atoms.
Most importantly, the strongest local correlations are associated with relative displacements of the ionic and CuO$_2$ layers, with correlation lengths similar to other important length scales in the system, notably the superconducting coherence length.
These displacements are not consistent with the direct effects expected due to the presence of interstitial oxygen atoms (required to dope the CuO$_2$ sheets) and therefore must arise from some other mechanism.
Our comprehensive approach, using neutron and x-ray diffuse scattering, \tdpdf/, and reverse Monte Carlo refinement, has enabled us to identify the relevant atomic displacements and to show that the correlations are likely unrelated to doping-induced disorder.
This implies that similar effects might be universally present in the cuprates and play an important role in their phenomenology.
We expect that the methodology developed here will be valuable for comparative studies of correlated bulk inhomogeneity in other cuprates, as well as for a wide range of materials with nanoscale correlations.

\begin{acknowledgments}
    We thank D.~Robinson and S.~Rosenkranz for assistance with x-ray scattering experiments.
    Work at the University of Minnesota was funded by the US Department of Energy through the University of Minnesota Center for Quantum Materials, under grant number DE-SC-0016371.
    Work at Argonne was supported by the U.S.
    Department of Energy, Office of Science, Basic Energy Sciences, Materials Sciences and Engineering Division.
    A portion of this research used resources at the Spallation Neutron Source, a DOE Office of Science User Facility operated by the Oak Ridge National Laboratory.
    This research used resources of the Advanced Photon Source, a US Department of Energy (DOE) Office of Science User Facility operated for the DOE Office of Science by Argonne National Laboratory under contract number DE-AC02-06CH11357.
    D.P.~acknowledges support from the Croatian Science Foundation through grant number UIP-2020-02-9494.
\end{acknowledgments}

\bibliography{main}

\end{document}